\documentstyle[12pt,amsfonts]{article}
\def\one{1\hskip-.37em 1}

\def\half{\textstyle{\frac{1}{2}}}

\def\H{{\cal H}}

\def\D{{\cal D}}
\def\t{\textstyle}
\def\s{\hskip.013cm}

\def\E{{\rm I}\hskip-.2em{\rm E}}
\def\ra{\rightarrow}
\def\tint{{\textstyle\int}}
\def\d{\partial}
\def\o{\overline}

\def\b{\begin{eqnarray*}}     
\def\e{\end{eqnarray*}}       
\def\bn{\begin{eqnarray}}     
\def\en{\end{eqnarray}}       
\def\<{\langle}
\def\>{\rangle}

\def\no{\nonumber}

\def\{{\lbrace}
\def\}{\rbrace}
\def\hg{{\hat g}}
\def\hp{{\hat\pi}}
\def\d{\delta}
\begin{document}

\date{}

\title{An Affinity for Affine Quantum Gravity}
\author{John R. Klauder\footnote{Electronic mail: john.klauder@gmail.com}\\
Department of Physics and \\Department of Mathematics\\
University of Florida\\
Gainesville, FL  32611\footnote{Presented at the Conference ``Gauge Fields. Yesterday, Today, and Tomorrow'',
Moscow, Russia, January 19-24, 2010, in honor of the 70th birthday of A. A. Slavnov.}}

\maketitle

\begin{abstract}

The main principle of affine quantum gravity is the strict positivity of the matrix $\{\hg_{ab}(x)\}$ composed of the spatial components of the local metric operator. Canonical commutation relations are incompatible with this principle, and they can be replaced by noncanonical, affine commutation relations. Due to the partial second-class nature of the quantum gravitational constraints, it is advantageous to use the  projection operator method, which treats all quantum constraints on an equal footing. Using this method, enforcement of regularized versions of the gravitational constraint operators  is formulated quite naturally as a novel and relatively well-defined functional integral involving only the same set of variables that appears in the usual classical formulation.  Although perturbatively nonrenormalizable, gravity may possibly be understood nonperturbatively from a hard-core perspective that has proved valuable for specialized models.
\end{abstract}

\section{Introduction}
Despite its difficulty, quantization of the gravitational field has attracted considerable attention because of its fundamental importance. Currently favored approaches
include string theory and loops, both important schemes; see, e.g., \cite{str,ash}.
A comparatively new effort is the {\it affine quantum gravity program}.  Although the principles involved are conservative and fairly natural, this program nevertheless involves a somewhat unconventional approach when compared with more traditional techniques. This article offers an overview of the affine quantum gravity  program; detailed discussions of this program appear in \cite{kla1, kla2, kla3}.

\subsubsection*{Basic principles of affine quantum gravity}

The program of affine quantum gravity is founded on {\it four basic
principles} which we briefly review here. {\it First}, like the corresponding classical variables, the 6 components of the spatial metric field operators $\hg_{ab}(x)\,[=\hg_{ba}(x)]$, $a,b=1,2,3$,  form a {\it positive-definite $3\!\times\! 3$ matrix for all $x$}. {\it Second}, to ensure self-adjoint kinematical variables when smeared, it is necessary to adopt the {\it affine commutation relations} (with $\hbar=1$)
\bn &&[\hp^a_b(x),\hp^c_d(y)]=i\half[\d^c_b\,\hp^a_d(x)-\d^a_d\,\hp^c_b(x)]\,\d(x,y)\;,\no\\
  && \hskip-.08cm[\hg_{ab}(x),\hp^c_d(y)]=i\half[\d^c_a\,\hg_{db}(x)+\d^c_b\,\hg_{ad}(x)]\,\d(x,y)\;,\\
  && \hskip-.18cm[\hg_{ab}(x),\hg_{cd}(y)]=0\;\no  \label{e1}\en
between the metric and the 9 components of the mixed-index momentum field operator $\hp^a_b(x)$, the
quantum version of the classical variable $\pi^a_b(x)\equiv \pi^{ac}(x)\s g_{cb}(x)$; these commutation
relations are direct transcriptions of Poisson brackets for the classical fields $g_{ab}(x)$ and $\pi^c_d(x)$. The affine commutation relations are like {\it current commutation relations} and their representations are
quite different from those for
canonical communtation relations; indeed, the present program is called `affine quantum gravity' because these commutation relations are analogous to
the Lie algebra belonging to affine transformations $(A,\s b)$, where $x\ra x'=A\s x+b$, $x,\s x'$, and $b$ are $n$-vectors, and $A$ are real, invertible $n\times n$ matrices. {\it Third}, the principle of {\it quantization first and reduce second}, favored by Dirac, requires that the basic fields $\hg_{ab}$ and $\hp^c_d$ are initially realized by an {\it ultralocal representation}, which is explained below. {\it Fourth}, and last, introduction and enforcement of the gravitational constraints not only leads to the physical Hilbert space but it has the added virtue that all vestiges of the temporary ultralocal operator representation are replaced by physically acceptable alternatives.
In attacking these basic issues full use of {\it coherent state methods} and the {\it projection operator method} for constrained system quantization is made.

\subsubsection*{Affine coherent states}

The affine coherent states are defined (for $\hbar=1$) by
  \bn |\pi,g\>\equiv e^{\t i\tint \pi^{ab}\hg_{ab}\,d^3\!x}\,e^{\t-i\tint\gamma^a_b\hp^b_a\,d^3\!x}\,|\eta\> \label{e2}\en
for general, smooth, $c$-number fields $\pi^{ab}\,[=\pi^{ba}]$ and $\gamma^c_d$ of compact support, and the
fiducial vector $|\eta\>$ is chosen so that the coherent state overlap functional becomes
 \bn  &&\hskip-.5cm\<\pi'',g''|\pi',g'\>= \exp\bigg(\!-\!2\int b(x)\,d^3\!x\, \no\\
  &&\hskip.1cm\times\ln\bigg\{  \frac{
\det\{\half[g''^{kl}(x) +g'^{kl}(x)]+i\half b(x)^{-1}[\pi''^{kl}(x)-
\pi'^{kl}(x)]\}} {(\det[g''^{kl}(x)])^{1/2}\,(\det[g'^{kl}(x)])^{1/2}}
\bigg\}\bigg) \;.\label{e3}\en
Observe that the matrices $\gamma''$ and $\gamma'$ do {\it not} explicitly appear in (\ref{e3}); the
 choice of $|\eta\>$ is such that each $\gamma=\{\gamma^a_b\}$ has  been replaced by $g=\{g_{ab}\}$, where
  \bn  g_{ab}(x)\equiv [e^{\t\gamma(x)/2}]_a^c\,\<\eta|\hg_{cd}(x)|\eta\>\,[e^{\t\gamma(x)^T/2}]_b^d \;.\en
Note that the functional expression in (\ref{e3}) is ultralocal, i.e., specifically of the form
  \bn  \exp\{-\tint b(x)\,d^3\!x\,L[\pi''(x),g''(x);\pi'(x),g'(x)]\,\}\;, \en
 and thus, {\it by design}, there are no correlations between spatially separated field values, a neutral position adopted towards correlations before any constraints are introduced. On invariance  grounds, (\ref{e3}) necessarily involves a {\it scalar density} $b(x)$, $0< b(x)<\infty$, for all $x$; this arbitrary and nondynamical auxiliary function $b(x)$ will disappear when the gravitational constraints are fully enforced, at which point proper field correlations will arise; see below. In addition, note that the coherent state overlap functional is {\it invariant} under general spatial coordinate transformations. Finally, we emphasize that the expression $\<\pi'',g''|\pi',g'\>$ is a {\it continuous functional of positive type} and thus may be used as a {\it reproducing kernel} to define a {\it reproducing kernel Hilbert space}
(see \cite{mesh}) composed of continuous phase-space functionals $\psi(\pi,g)$ on which the initial, ultralocal representation of the affine field operators acts in a natural fashion.

\subsubsection*{Functional integral representation}
 A functional integral formulation has been developed \cite{kla2} that, in effect, {\it within a single formula} captures the essence of {\it all four of the basic principles} described above. This ``Master Formula'' takes the form
\bn  && \<\pi'',g''|\s\E\s|\pi',g'\>  \no\\
 &&\hskip1cm=\lim_{\nu\ra\infty}{\o{\cal N}}_\nu\s\int
e^{\t-i\tint[g_{ab}{\dot\pi}^{ab}+N^aH_a+NH]\,d^3\!x\,dt}\no\\
  &&\hskip1.5cm\times\exp\{-(1/2\nu)\tint[b(x)^{-1}g_{ab}g_{cd}
{\dot\pi}^{bc}{\dot\pi}^{da}+b(x)g^{ab}g^{cd}{\dot g}_{bc}{\dot g}_{da}]\,
d^3\!x\,dt\}\no\\
  &&\hskip2cm\times[\Pi_{x,t}\,\Pi_{a\le b}\,d\pi^{ab}(x,t)\,
dg_{ab}(x,t)]\,\D R(N^a,N)\;. \label{e8} \en
Let us explain the meaning of (\ref{e8}).

As an initial remark, let us
artificially set $H_a=H=0$, and use the fact that $\int{\cal D}R(N^a,N)=1$. Then the result is that $\E=\one$, and the remaining functional integral yields the coherent state overlap $\<\pi'',g''|\pi',g'\>$ as given in (\ref{e3}). This is the state of affairs {\it before} the constraints are imposed, and remarks below regarding the properties of the functional integral on the right-hand side of (\ref{e8}) apply in this case as well. We next turn to the full content of (\ref{e8}).

The expression $\<\pi'',g''|\s\E\s|\pi',g'\>$  denotes the coherent state matrix element of a projection operator $\E$ which projects onto a subspace of the original Hilbert space on which the quantum constraints are fulfilled in a
regularized fashion. Furthermore, the expression $\<\pi'',g''|\s\E\s|\pi',g'\>$ is another continuous functional of positive type that can be used as a reproducing kernel to generate the reproducing kernel physical Hilbert space on which the quantum constraints are fulfilled in a regularized manner. The right-hand side of equation (\ref{e8}) denotes an essentially well-defined functional integral over fields $\pi^{ab}(x,t)$ and $g_{ab}(x,t)$, $0<t<T$, designed to calculate this important reproducing kernel for the regularized physical Hilbert space and which entails functional arguments defined by their smooth initial values $\pi^{ab}(x,0)=\pi'^{ab}(x)$ and $g_{ab}(x,0)=g'_{ab}(x)$ as well as their smooth final values $\pi^{ab}(x,T)=\pi''^{ab}(x)$ and $g_{ab}(x,T)=g''_{ab}(x)$, for all $x$ and all $a,b$. Up to a surface term, the phase factor in the functional integral represents the canonical action for general relativity, and specifically $N^a$ and $N$ denote Lagrange multiplier fields (classically interpreted as the shift and lapse), while $H_a$ and $H$ denote phase-space symbols (since $\hbar\ne0$) associated with the quantum diffeomorphism and Hamiltonian constraint field operators, respectively. The $\nu$-dependent factor in the integrand formally tends to unity in the limit $\nu\ra\infty$; but prior to that limit, the given expression {\it regularizes and essentially gives genuine meaning} to the heuristic, formal functional integral that would otherwise arise if such a factor were missing altogether \cite{kla2}. The functional form of the given regularizing factor ensures that the metric variables of integration {\it strictly fulfill} the positive-definite domain requirement. The given form, and in particular the need for the nondynamical, nonvanishing, arbitrarily chosen scalar density $b(x)$, is very welcome  since this form---{\it and quite possibly only this form}---leads to a reproducing kernel Hilbert space for gravity having the needed infinite dimensionality; a seemingly natural alternative \cite{kla5} using $\sqrt{\det[g_{ab}(x)]}$ in place of
$b(x)$ fails to lead to a reproducing kernel Hilbert space with the required dimensionality \cite{wat2}. The choice of $b(x)$ determines a specific ultralocal representation for the basic affine field variables, but this unphysical and temporary representation {\it disappears} after the gravitational constraints are fully enforced (as soluble examples explicitly demonstrate \cite{kla3}). The integration over the Lagrange multiplier fields ($N^a$ and $N$) involves a {\it specific measure} $R(N^a,N)$ (described in \cite{kla6}), which is normalized such that $\tint\D R(N^a,N)=1$. This measure is designed to enforce (a regularized version of) the
{\it quantum constraints$\s$}; it is manifestly {\bf not} chosen to enforce the classical constraints, even in a regularized form. The consequences of this choice are {\it profound} in that no gauge fixing is needed, no ghosts are required, no Dirac brackets are necessary, etc. In short, {\it no auxiliary structure of any kind is introduced}. (These facts are general properties of the projection operator method of dealing with constraints \cite{kla6,sch} and are not limited to gravity. A sketch of this method appears below.)

It is fundamentally important to make clear how Eq.~(\ref{e8}) was derived and how it is to be used \cite{kla2}. The left-hand side of (\ref{e8}) is an abstract operator construct in its entirety that came {\it first} and corresponds to one of the basic expressions one would like to calculate. The functional integral on the right-hand side of (\ref{e8}) came {\it second} and is a valid representation of the desired expression; its validity derives from the fact that the affine coherent state representation enjoys a complex polarization that is used to formulate a kind of Feynman-Kac realization of the coherent state matrix elements of the regularized projection operator \cite{kla2}.
 However, the final goal is to turn that order around and to use the functional integral to {\it define and evaluate} (at least approximately) the desired operator-defined expression on the left-hand side. In no way should it be thought that the functional integral (\ref{e8}) was ``simply postulated as a guess as how one might represent the proper expression''.

 A major goal in the general analysis of (\ref{e8}) involves reducing the regularization imposed on the quantum constraints to its appropriate minimum value, and, in particular, for constraint operators that are partially second class, such as those of gravity, the proper minimum of the regularization parameter is {\it non\/}zero;
 see below. Achieving this minimization involves {\it fundamental changes} of the representation of the basic kinematical operators, which, as models show \cite{kla3}, are so significant that any unphysical aspect of the original, ultralocal representation disappears completely. When the appropriate minimum regularization is achieved, then the quantum constraints are properly satisfied. The result is the reproducing kernel for the physical Hilbert space, which then permits a variety of physical questions to be studied.

 We next offer some additional details.

\section{Quantum Constraints and their Treatment}
The quantum gravitational constraints, $\H_a(x)$, $a=1,2,3$, and $\H(x)$, formally satisfy the commutation relations
 \bn &&[\H_a(x),\H_b(y)]=i\s\half\s[\delta_{,a}(x,y)\s\H_b(y)+\delta_{,b}(x,y)\s\H_a(x)]\;,\no\\\
  &&\hskip.15cm[\H_a(x),\H(y)]=i\s\delta_{,a}(x,y)\s\H(y) \;,\\
  &&\hskip.31cm[\H(x),\H(y)]=i\s\half\s\delta_{,a}(x,y)\s[\s g^{ab}(x)\s\H_b(x)+\H_b(x)\s g^{ab}(x) \no\\
&&\hskip3.6cm +g^{ab}(y)\s\H_b(y)+\H_b(y)\s g^{ab}(y)\s] \;. \no  \en
Following Dirac, we first suppose that $\H_a(x)\s|\psi\>_{phys}=0$ and $\H(x)\s|\psi\>_{phys}=0$ for all $x$ and $a$, where $|\psi\>_{phys}$ denotes a vector in the proposed physical Hilbert space ${\frak H}_{phys}$. However, these conditions are {\it incompatible} since, generally, $[\H(x),\H(y)]\s|\psi\>_{phys}\ne0$ because $[\H_b(x),g^{ab}(x)]\ne0$ and $g^{ab}(x)\s|\psi\>_{phys}\not\in{\frak H}_{phys}$, even when smeared. This means that the quantum gravity constraints are {\it partially second class}. While others may resist this conclusion, we accept it for what it is.

One advantage of the projection operator method is that it treats first- and second-class constraints on an {\it equal footing$\s$}; see \cite{kla6,sch}. The essence of the projection operator method is the following. If $\{\Phi_a\}$ denotes a set of self-adjoint quantum constraint operators, then
  \bn  \E=\E(\!\!(\Sigma\s\Phi_a^2\le\s\delta(\hbar)^2\s)\!\!) =\int {\sf T}\s e^{\t-i\tint\lambda^a(t)\s\Phi_a\,dt}\,\D R(\lambda)\;, \label{e10}\en
  in which ${\sf T}$ enforces time ordering,
denotes a projection operator onto a regularized physical Hilbert space, ${\frak{H}}_{phys}\equiv\E\s\frak{H} $,
where $\frak{H}$ denotes the original Hilbert space before the constraints are imposed.
 It is noteworthy that there
is a {\it universal form} for the weak measure $R$ \cite{kla6} that depends only on the
number of constraints, the
time interval involved, and the regularization parameter $\delta(\hbar)^2$; $R$ does {\it not} depend in any
way on the constraint operators themselves! Sometimes, just by reducing the regularization parameter $\delta(\hbar)^2$ to its appropriate size, the proper physical Hilbert space arises. Thus, e.g., if $\Sigma\s\Phi_a^2=J_1^2+J_2^2+J_3^2$, the Casimir operator of $su(2)$, then $0\le\delta(\hbar)^2<3\hbar^2/4$ works for this first class example. If $\Sigma\s\Phi_a^2=P^2+Q^2$, where $[Q,P]=i\hbar\one$, then $\hbar\le\delta(\hbar)^2<3\hbar$ covers this second class example. Sometimes, one needs to take the limit when
$\delta\ra0$. The example $\Sigma\s\Phi_a^2=Q^2$ involves
a case where $\Sigma\Phi^2_a=0$ lies in the continuous spectrum. To deal with this case it is appropriate to introduce
  \bn \<\<p'',q''|p',q'\>\>\equiv \lim_{\delta\ra0}\<p'',q''|\s\E\s|p',q'\>/\<\eta|\s\E\s|\eta\>\;, \en
where $\{|p,q\>\}$ are traditional coherent states, as a reproducing kernel for the physical Hilbert space in which ``$Q=0$'' holds. It is interesting to observe that the projection operator for {\it reducible} constraints, e.g., $\E(Q^2+Q^2\le\delta^2)$, or for {\it irregular} constraints, $\E(Q^{2\Omega}\le\delta^2)$, $0<\Omega\ne1$, leads to the {\it same reproducing kernel} that arose from the case $\E(Q^2\le\delta^2)$. No gauge fixing is ever needed, and thus no global consistency conditions arise that may be violated; see, e.g., \cite{litkla}.
Other cases may be more involved but the principles are similar. The time-ordered integral representation for $\E$ given in (\ref{e10}) is useful in path integral representations, and this application explains the origin of $R(N^a,N)$ in (\ref{e8}).

\section{Nonrenormalizability and Symbols}
Viewed perturbatively, gravity is nonrenormalizable. However, the (nonperturbative) {\it hard-core picture of nonrenormalizability} \cite{kla11,book} holds that the nonlinearities in such theories are so strong that, from a functional integral point of view, a nonzero set of functional histories that were allowed in the support of the linear theory is now forbidden by the nonlinear interaction.
\subsubsection*{Elementary example of hard-core behavior}
An elementary example that illustrates the basic concepts is the following. Consider a one-dimensional harmonic
oscillator with the Euclidean-time action functional
  \bn I_0\equiv\int_0^T \{\half[{\dot x}(t)^2+x(t)^2]\s\}\,dt \en
  defined for a set of functions $W_0\equiv\{x(t): \tint_0^T[\s{\dot x}^2+x^2]\s dt<\infty\}$. Observe that the set $W_0$  includes many functions for which $\tint_0^T x^{-4}\s dt=\infty$. As a consequence, the
  Euclidean-time action functional
   \bn I_g\equiv\int_0^T\{\half[{\dot x}(t)^2+x(t)^2] +g\s x(t)^{-4}\s\}\,dt \;,\en
   for $g>0$, defined for the set $W_g\equiv\{x(t): \tint_0^T[\s{\dot x}^2+x^2+x^{-4}]\s dt<\infty\}$ does
   {\it not} pass to the set $W_0$ as $g\ra0$. Stated otherwise -- now using a real time formulation -- the set of solutions of the interacting
   model with $g>0$ does {\it not} pass to the set of solutions of the free theory as $g\ra0$. Instead,
   the set of solutions pass to those of a {\it pseudofree} theory for which $x(t)=0$ is forbidden!

   This change in character of the classical theory has its image in the quantum theory as well. If
   $\{h_n(x)\}_{n=0}^\infty$ denotes the set of Hermite functions (eigenfunctions of the free harmonic oscillator), then
   the free quantum theory is determined  by the Euclidean-time propagator
        \bn {\cal N}_0\int_{x(0)=x'}^{x(T)=x''} e^{\t-I_0}\;{\cal D}x=\sum_{n=0}^\infty h_n(x'')\s h_n(x')\, e^{\t-(n+1/2)\s T}\;.\en
   In contrast, the Euclidean-time propagator for the quantum pseudofree theory is determined by
   [using $\theta(y)=1$, for $y>0$, and $\theta(y)=0$, for $y<0$]
      \bn &&\lim_{g\ra0} {\cal N}_g\int_{x(0)=x'}^{x(T)=x''} e^{\t-I_g}\,{\cal D}x=\theta(x''\s x')\sum_{n=0}^\infty\no\\
      &&\hskip2cm\times\{\s h_n(x'')[\s h_n(x')-h_n(-x')\s]\s\}\,e^{\t-(n+1/2)\s T} \en
      due to the hard-core nature of the potential that remains even after $g\ra0$. In brief, as $g\ra0$, the
      interacting theory is continuously connected to the pseudofree theory and not to the usual free
      theory; if a perturbation analysis of the interacting model is made, it must be made about
      the pseudofree theory and not about the free theory!
\subsubsection*{Hard-core behavior for field theories}
Various, highly specialized, nonrenormalizable quantum field theory models exhibit entirely analogous hard-core behavior, and nevertheless possess suitable nonperturbative solutions \cite{book}. It is believed that gravity and also $\phi^4$ field theories in high enough spacetime dimensions can be understood in similar terms. A computer study to analyze the $\phi^4$ theory has begun, and there is hope to clarify that particular theory. Any progress in the scalar field case could strengthen a similar argument for the gravitational case as well.

The expectation that nonrenormalizable, self-interacting scalar fields will exhibit hard-core behavior follows from a so-called multiplicative inequality \cite{lady,book}. In particular, for smooth
functions $\phi(x)$, $x\in{\mathbb{R}}^n,$ it follows that
   \bn \{\tint \phi(x)^4\,d^n\!x\s\}^{1/2}\le C\s\tint\{\s[\s{\nabla\phi}(x)\s]^2+m^2\s\phi(x)^2\s\}\,d^n\!x\;,\en
   where for $n\le 4$ (the renormalizable models) one may choose $C=4/3$, while for $n\ge5$ (the nonrenormalizable models) one must choose $C=\infty$ meaning, in the latter case,
   that there are field functions, e.g.,
   $\phi_{singular}(x)=|x|^{-p}\s \exp(-x^2)$, with $n/4\le p<n/2-1$, for which the left side diverges
   while the right side is finite. This is exactly the signal that the interaction acts at the
   classical level partially as a hard core, and it is not too much to expect that the quantum theory
   would also reflect that fact as well. Additional arguments favoring a hard-core understanding
   for $\phi^4$ models in five and more spacetime dimensions appear in \cite{kla17}.

Evidence from soluble examples points to the appearance of a nontraditional and nonclassical (proportional to $\hbar^2$) counterterm in the functional integral representing the irremovable effects of the hard core.
For the proposed quantization of gravity, these counterterms would have an important role to play in conjunction with the symbols representing the diffeomorphism and Hamiltonian constraints in the functional integral since for them $\hbar\ne0$ as well. In brief, the form taken by the symbols $H_a$ and $H$ in (\ref{e8}) is closely related to a proper understanding of how to handle the perturbative nonrenormalizability and the concomitant hard-core nature of the overall  theory. These are clearly difficult issues, but it is equally clear that they may be illuminated by studies of other nonrenormalizable models such as $\phi^4$ in five and more spacetime dimensions.

\section{Classical Limit}
Suppose one starts with a classical theory, quantizes it, and then takes the classical limit. It seems obvious that the classical theory obtained at the end should coincide with the classical theory one started with. However, there are counterexamples to this simple wisdom! For example, the $\phi^4$ theory in five spacetime dimensions has a {\it nontrivial} classical behavior. But, if one quantizes it as the continuum  limit of a natural lattice formulation, allowing for mass, field strength, and coupling constant renormalization, the result is a free (or generalized free) quantum theory whose classical limit is also free and thus differs from the original theory \cite{fro}. This unsatisfactory behavior is yet another facet  of the nonrenormalizability puzzle. However, those nonrenormalizable models for which the quantum hard-core behavior has been accounted for do have satisfactory classical limits \cite{book}. The conjectured hard-core nature of $\phi^4$ models is under present investigation, and it is anticipated that a proper classical limit should arise. It is further conjectured that a favorable consequence of clarifying and including the hard-core behavior in gravity will ensure that the resultant quantum theory enjoys the correct classical limit.

An additional remark may be useful. It is a frequent misconception that passage to the classical limit requires that the parameter $\hbar\ra0$. To argue against this view, recall that the macroscopic world we know and describe so well by classical mechanics is the same real world in which $\hbar\ne0$. In point of fact, classical and quantum formalisms must {\it coexist}, and this coexistence is very well expressed with the help of coherent states. It is characteristic of coherent state formalisms that classical and quantum ``generators'', loosely speaking, are related to each other through the {\it weak correspondence principle} \cite{corr}. In the case of the gravitational field, prior to the introduction of constraints, this connection takes the general form
  \bn  \<\pi,g|\s{\cal W}\s|\pi,g\>=W(\pi,g) \;,  \en
where $\cal W$ denotes a quantum generator and $W(\pi,g)$ the corresponding classical generator (which is generally a ``symbol'' still since $\hbar\ne0$ ). The simplest examples of this kind are given by $\<\pi,g|\s\hg_{ab}(x)\s|\pi,g\>=g_{ab}(x)$ and $\<\pi,g|\s\hp_a^b(x)\s|\pi,g\>=\pi^{bc}(x)g_{ca}(x)\equiv \pi_a^b(x)$. Moreover, these two examples also establish that the {\it physical meaning of the $c$-number labels
is that of  mean values} of the respective quantum field operators in the affine coherent states.

In soluble models where the appropriate classical limit has been obtained \cite{book}, coherent state methods were heavily used. It is expected that they will prove equally useful in the case of gravity.

\section{Going Beyond the Ultralocal\\ Representation}
We started our discussion by choosing an ultralocal representation of the basic affine quantum field operators.
Before the constraints were introduced, an ultralocal representation is the proper choice because all the
proper spatial connections are contained in the constraints themselves. Moreover, the chosen ultralocal representation is based on an extremal weight vector of the underlying affine algebra, which has the virtue of
leading to affine coherent states that fulfill a complex polarization condition enabling us to obtain
a fairly well defined functional integral representation for coherent state matrix elements of the
regularized projection operator. To complete the story, one only needs to eliminate the regularizations!
Of course, this is an enormous task. But it should not be regarded as impossible because there is a model
problem in which just that issue has been successfully dealt with. In \cite{kla3} the quantization of a
free field of mass $m$ (among other examples) was discussed starting with a reparametrization invariant formulation. In particular, by elevating the time variable to a dynamical one, the original dynamics is transformed to the imposition of a constraint. Thus, in the constrained form, the Hamiltonian vanishes, and
the choice of the original representation of the field operators is taken as an ultralocal one. Subsequent imposition of the constraint---by the projection operator method---not only eliminated the ultralocal
representation but allowed us to focus the final reproducing kernel for the physical Hilbert space on any
value of the mass parameter $m$ one desired! It is the kind of procedures used for this relatively simple example
of free field quantization that we have in mind to be used to transform the original ultralocal representation
 of the quantum gravity story into its final and physically relevant version.

 \section{Dedication} I am pleased to dedicate this article to Andrei Alekseevich Slavnov, a scholar and gentleman of the first rank. I hope he enjoys many more years of productive research!

\section{Acknowledgements}
Thanks are extended to the Center for Applied Mathematics of the Mathematics Department, University of Florida,
for partial travel support to attend the ``Slavnov70'' conference.

\end{document}